\documentclass[11pt]{article}

\usepackage{amsmath,amssymb,amsfonts}
\usepackage{bm}
\usepackage{geometry}
\usepackage{hyperref}
\usepackage{mathtools}
\DeclareMathOperator{\sech}{sech}
\DeclareMathOperator{\csch}{csch}
\title{\bf The Nakamura Conjecture Revisited: Toda Molecules and Stationary Axisymmetric Gravity}

\author{
Takeshi Fukuyama\\[2mm]
Research Center for Nuclear Physics (RCNP), Osaka University,\\
Ibaraki, Osaka 567--0047, Japan
}

\date{}

\begin{document}

\maketitle

\begin{abstract}
We revisit the Nakamura conjecture, which relates the Tomimatsu–Sato solutions of stationary axisymmetric gravity to finite Toda molecules. While the conjecture has been established partially, its general rotating sector remains an open problem.

We show that the Toda determinants underlying the conjecture possess a natural weight grading. In particular, the two functions entering the Ernst potential have weights \(n^2\) and \(n^2-1\), and this grading extends systematically to shifted determinants labelled by partitions. In coordinates adapted to the Toda generators, each differentiation corresponds to adding one box to the associated Young diagram and increases the weight by one. The same integer \(n^2\) also appears in the zero-order term of the Nakamura bilinear operator, revealing a compatibility between the differential equation and the determinant grading.

The partition structure further explains the previously unresolved behavior of second derivatives. Repeated differentiation in one direction produces an internal sector and an external sector requiring only a one-step extension of the Wronskian hierarchy; the latter is reduced by a local three-term Plücker relation. Thus weight grading, Young-diagram growth, Wronskian enlargement, and Plücker reduction emerge as parts of a single determinant structure.

The unit weight relation \(n^2=(n^2-1)+1\) also singles out the elementary Toda seed as a natural third object, suggesting a possible route toward a genuine trilinear formulation. Although no trilinear closure is assumed here, the present construction reduces the remaining general-\(n\) Nakamura problem to definite determinant-minor identities and provides a structural framework in which such a formulation can be investigated.

\end{abstract}

\section{Introduction}
\label{sec:introduction}

Stationary and axisymmetric vacuum gravity provides an important
setting in which the nonlinear Einstein equations admit an
integrable reduction. Stationarity and axial symmetry imply two
commuting Killing directions, corresponding to time translation and
rotation about the symmetry axis. After these symmetry directions
are separated, the gravitational field depends on two coordinates,
which may be chosen as the usual prolate spheroidal coordinates
$(x,y)$. The vacuum Einstein equations can then be written in terms
of a complex Ernst potential \cite{Ernst}.

Among the exact solutions of this system, the Kerr solution \cite{Kerr1963} describes
the exterior field of a rotating black hole, while the
Tomimatsu--Sato (TS) solutions
\cite{TomimatsuSato1972,TomimatsuSato} form a discrete
generalization characterized by an integer deformation parameter
$\delta=n$. The case $n=1$ corresponds to Kerr, whereas higher
integer values describe the higher members of the TS family. In
addition to this discrete label, the solution contains the parameters
$p$ and $q$, conventionally normalized by
\begin{equation}
        p^2+q^2=1,
\label{pq}
\end{equation}
with $q$ controlling the rotational part of the solution. Thus the
TS family combines a continuous rotational parameter with the
discrete deformation index $n$.

The TS solutions were subsequently generalized by Yamazaki and Hori
to arbitrary integral values of the deformation parameter
\cite{YamazakiHori,Hori1978}. Their determinant representation was
later reformulated in an equivalent persymmetric-determinant form by
Vein \cite{Vein}. More recently, Melikyan
\cite{Melikyan2025,Melikyan2026} revisited the algebraic structure
of the Yamazaki--Hori solutions. The former work also established a
connection between the Yamazaki--Hori construction and a nonlinear
differential equation introduced by Cosgrove \cite{Cosgrove}. In
particular, a systematic factorization of the associated polynomials
was obtained, and in the nonrotating Weyl sector $(q=0)$ \cite{Weyl1917} the
corresponding reduced polynomials were identified with
$\tau$-functions of the semi-infinite two-dimensional Toda molecule.
These developments provide a modern algebraic perspective on the
determinant structure underlying the TS family.

A different, but closely related, development was initiated by
Nakamura, who discovered a remarkable relation between these
gravitational solutions and the Toda molecule
\cite{Toda1967,Nakamura,NakamuraOhta}. For the Pfaffian representation of the TS solutions,
Nakamura and Ohta explicitly verified the bilinear relations
through $n=6$: the cases $n=1,\ldots,4$ were treated explicitly,
while $n=5$ and $6$ were checked by computer algebra using
REDUCE3 \cite{NakamuraOhta}.  They identified the remaining
problem as an analytic proof for arbitrary $n$. The significance of the
discrete parameter $n$ is not merely technical. From the
gravitational point of view, it is not obvious why the TS deformation
should be organized by an integer $n=1,2,\ldots$. The field equations
themselves are differential equations in continuous spacetime
variables, and the origin of this discrete hierarchy is therefore
not transparent. The Toda-molecule representation offers a natural
structural interpretation: the same integer $n$ labels the size of
the finite Toda molecule, or equivalently the order of the determinant
generated from the Toda lattice. Thus the discrete deformation
parameter on the gravitational side is mapped to an intrinsically
discrete quantity on the integrable-system side. This observation
does not by itself explain or prove a quantization of the
gravitational solutions, but it suggests that their discrete
organization may have its origin in the finite-lattice structure of
the underlying Toda system.

This is one reason why a proof of the Nakamura conjecture would be of
interest beyond the construction of the TS solutions themselves. It
would establish a precise bridge between two theories with
independently rich structures: stationary axisymmetric gravity and
soliton integrable systems. Such a bridge may allow structures which
appear unexplained on one side to acquire a natural meaning on the
other. Conversely, gravitational properties may reveal new
constraints or interpretations of the corresponding Toda hierarchy.
The integer $n$ provides a simple example of this possibility, but
the broader motivation is to determine whether the determinant,
lattice, and algebraic structures of soliton theory can expose hidden
organizing principles of nonlinear gravitational solutions.

The emergence of such an intrinsically discrete structure within a
family of classical gravitational solutions may also be relevant from
the viewpoint of quantization. The integer $n$ is, of course, a
label of classical solutions and should not by itself be interpreted
as a quantum number. Nevertheless, if the correspondence with the
Toda molecule can be established at the structural level, its
determinant order, grading, and discrete lattice structure may provide
natural variables in which the quantization of this gravitational
sector can be reconsidered. This possibility provides an additional
motivation for clarifying the Nakamura conjecture beyond the
construction of individual exact solutions.

The essential content of the Nakamura conjecture may be summarized
as follows \cite{Nakamura,NakamuraOhta,FKY}. For a stationary
axisymmetric vacuum spacetime, the Ernst equation can be written in
the form
\begin{equation}
 (\xi\xi^*-1)\nabla^2\xi
 -2\xi^*\,\nabla\xi\cdot\nabla\xi=0 .
\label{ernst}
\end{equation}
Writing the Ernst potential of the $n$-th TS solution in terms of
the two TS functions $g_n$ and $f_n$ as
\begin{equation}
        \xi_n=\frac{g_n}{f_n},
\label{xirat}
\end{equation}
this nonlinear equation can be decomposed into two pairs of bilinear
relations. The first pair involves first Hirota derivatives,
\begin{equation}
\begin{split}
 D_x(g_n\cdot f_n-g_n^*\cdot f_n^*)&=0,\\
 D_y(g_n\cdot f_n+g_n^*\cdot f_n^*)&=0,
\end{split}
\label{bil1}
\end{equation}
whereas the second pair is
\begin{equation}
\begin{split}
 {\cal F}(g_n^*\cdot f_n)&=0,\\
 {\cal F}(g_n^*\cdot g_n+f_n^*\cdot f_n)&=0 .
\end{split}
\label{bil2}
\end{equation}
Here ${\cal F}$ is the second-order bilinear operator
\begin{equation}
 {\cal F}
 =
 (x^2-1)D_x^2+2x\partial_x
 +(y^2-1)D_y^2+2y\partial_y-2n^2 .
\label{Fop}
\end{equation}
These equations provide a convenient formulation of the Nakamura
problem.

The second ingredient of the conjecture is the Toda-molecule
construction of the functions $g_n$ and $f_n$. We briefly recall
the construction used in Ref.~\cite{FKY}. A Toda molecule is a
finite Toda system whose tau functions can be represented by
Wronskian-type determinants generated from a single function $\psi$.
For the present two-variable problem, introduce the commuting
differential operators
\begin{equation}
 L_{\pm}
 =
 (x^2-1)\partial_x
 \pm (y^2-1)\partial_y .
\label{Lpm}
\end{equation}
Starting from $\psi$, the $n$-point Toda tau function is constructed
as
\begin{equation}
 \tau_n
 =
 \det\left[
 L_+^{\,i-1}L_-^{\,j-1}\psi
 \right]_{i,j=1}^{n}.
\label{tauintro}
\end{equation}
Thus $\psi$ is the basic generating function from which the entire
Toda determinant hierarchy is produced by successive actions of
$L_+$ and $L_-$.

Nakamura's observation was that the TS functions can be obtained from
this Toda construction by choosing
\begin{equation}
        \psi=px-iqy ,
\label{psi}
\end{equation}
with $p^2+q^2=1$. More precisely, $g_n$ is identified with the
$n$-point Toda tau function,
\begin{equation}
        g_n=\tau_n ,
\label{gntau}
\end{equation}
whereas $f_n$ is obtained from the corresponding $(n-1)$-point
determinant with the generating function shifted by $L_+L_-$,
\begin{equation}
        f_n=
        \left.
        \tau_{n-1}
        \right|_{\psi\rightarrow L_+L_-\psi}.
\label{fntau}
\end{equation}
The Nakamura conjecture therefore states that these particular Toda
determinants satisfy the bilinear relations above and hence generate
the Ernst potential of the $n$-th TS solution.

The conjecture was investigated analytically by Fukuyama, Kamimura
and Yu (FKY) \cite{FKY}. By embedding $g_n$, $f_n$, and their
complex conjugates into a common enlarged Wronskian (reintroduced
in Sec.~6), the first pair of bilinear equations was reduced to
Pfaffian identities and proved for generic rotation. The second pair,
containing the second-order operator ${\cal F}$, was established
there in the nonrotating case $q=0$. A complementary approach was
developed by Imai and Fukuyama \cite{ImaiFukuyama}, who used Aitken
acceleration to establish the connection between Toda-molecule
solutions and the stationary axisymmetric Einstein equation from a
different viewpoint. Further progress on the generic rotating
problem was made by Fukuyama and Koizumi \cite{FK2011}, but a proof
of the complete second-order system for arbitrary $n$ and generic
rotation has remained open.

The recent work of Melikyan \cite{Melikyan2025,Melikyan2026}
provides important additional information on the algebraic structure
of the Yamazaki--Hori solutions and their relation to Toda
$\tau$-functions. The problem considered here is complementary:
we focus on the differential and minor structure required by the
second-order Nakamura equations in the generic rotating sector.
In particular, we ask how the Toda determinant hierarchy behaves
under the successive derivatives appearing in ${\cal F}$.

The purpose of the present paper is therefore to reconsider this
unresolved second-order problem directly from the structure of the
Toda determinants. Our first observation is that the variables
$(x,y)$ are not the most transparent coordinates for this purpose.
We introduce new variables $(\xi,\eta)$ through
\begin{equation}
        x=\coth\xi,\qquad y=\tanh\eta .
\label{coord}
\end{equation}
In these variables the differential operators which generate the
Toda Wronskians become constant-coefficient derivatives, while the
first-derivative terms appearing explicitly in ${\cal F}$ disappear.
This makes it possible to separate more clearly the differential
structure of the Nakamura equation from the algebraic structure of
the Toda determinants.

The same change of variables reveals a simple grading hidden in the
Toda functions. Their hyperbolic factors can be classified by the
total powers of $(\sinh\xi)^{-1}$ and $(\cosh\eta)^{-1}$. A
determinant counting argument gives the exact weights $n^2$ and
$n^2-1$ for $g_n$ and $f_n$. We then introduce, in Sec.~5, a
two-sided shifted determinant hierarchy $\tau_m^{[r,s]}$, which
extends the ordinary common-shift Toda family and resolves its
derivatives by partitions. The weight increases by exactly one for
each added box.

\begin{figure}[t]
\centering
\includegraphics[width=0.92\textwidth]{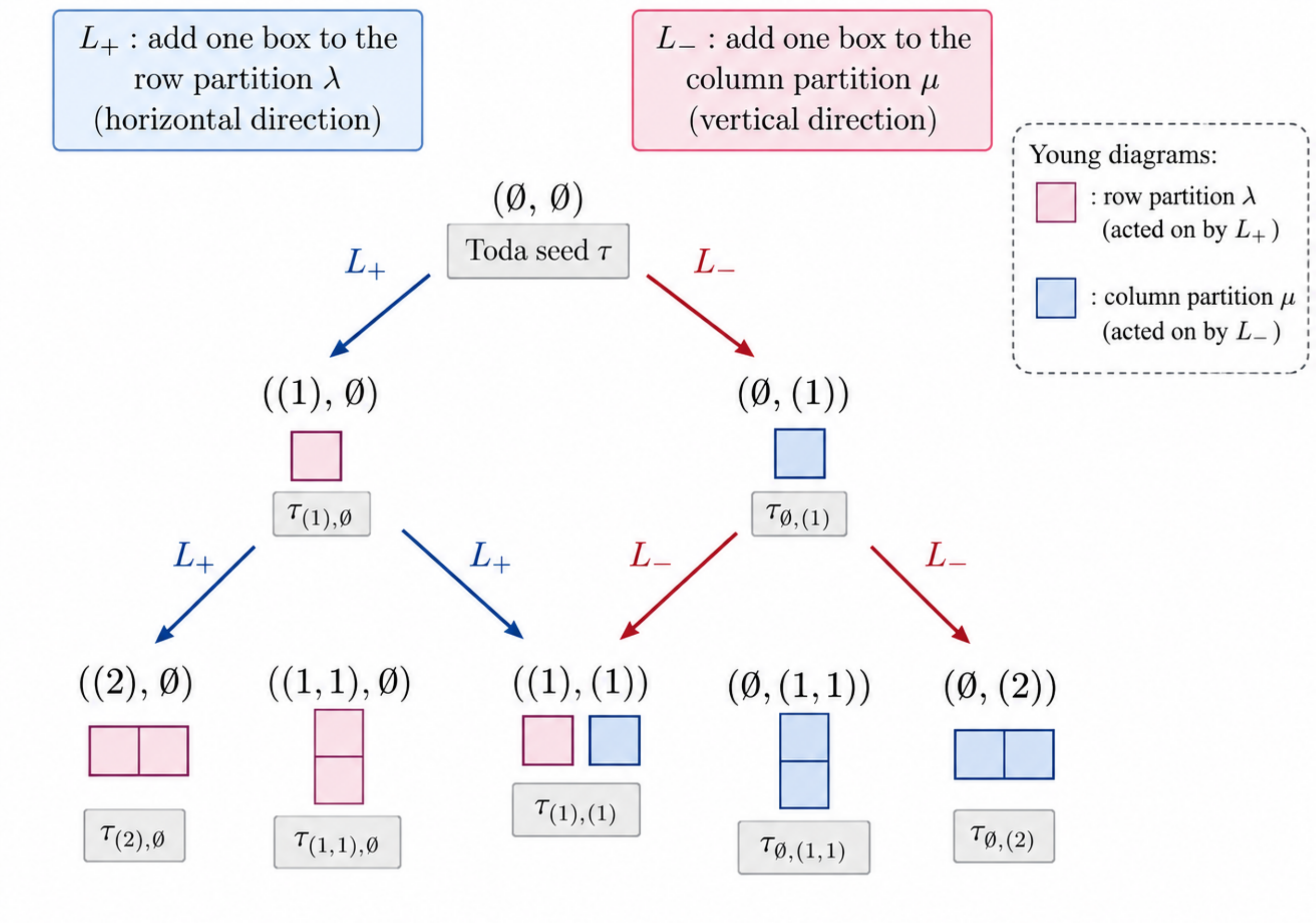}
\caption{Partition structure generated by $L_+$ and $L_-$.
Starting from the empty partition, successive actions of $L_+$
and $L_-$ generate the row and column Young sectors, respectively.
At second order the pure derivatives give the partitions $(2)$
and $(1,1)$, while the mixed derivative gives $((1),(1))$.}
\label{fig:young}
\end{figure}

This partition structure gives a direct explanation of the second
Wronskian derivatives. The structure relevant below is summarized
schematically in Fig.~\ref{fig:young}. Each action of $L_+$
($L_-$) adds one box to the row (column) partition. Thus, at second
order, the pure derivatives generate the two Young sectors $(2)$
and $(1,1)$, whereas the mixed derivative generates the sector
$((1),(1))$.

Mixed differentiation produces the sector $((1),(1))$ and remains
inside the $(n+2)\times(n+2)$ enlarged determinant $D$ introduced
in Sec.~6. Pure differentiation splits into $(1,1)$ and $(2)$
sectors. The $(1,1)$ contribution stays inside the original
Wronskian range, whereas the $(2)$ contribution reaches one step
beyond it. The two external terms $E_+$ and $E_-$, which arise
explicitly from the pure second derivatives in
Eqs.~(\ref{Hpp}) and (\ref{Gmm}), are precisely the row and column
$(2)$ sectors. Furthermore, their products with the relevant
Nakamura pair $(g_n^*,f_n)$ are reduced by the same local
three-term Pl\"ucker relation. The one-step Wronskian enlargement
$D_{n+2}\to\widehat D_{n+3}$ and the Pl\"ucker reduction are
therefore organized by a common determinant geometry.

The present paper does not assume a trilinear closure. Hirota's
bilinear formalism and its multilinear extensions remain useful
background \cite{Hirota,HirotaBook,GRH,YTSF1}, but the main purpose
here is narrower: to identify a general-$n$ determinant principle
which can be checked independently of any proposed three-slot
equation. Whether a genuine trilinear structure emerges from this
partition-resolved hierarchy is left as a separate question.

The paper is organized as follows. In Sec.~2 we introduce coordinates
adapted to the Toda generating operators, rewrite the Nakamura
operator in these variables, and develop the corresponding
hyperbolic representation of the Toda building blocks. Section~3
establishes the weight grading of the Toda determinants, while
Sec.~4 illustrates this structure in the nonrotating sector and in
the explicit rotating $n=2$ solution. In Sec.~5 we embed the
Nakamura pair into a shifted Toda determinant lattice and introduce
the partition-resolved derivative structure, in which the actions
of $L_+$ and $L_-$ correspond to adding boxes to row and column
Young diagrams. Section~6 applies this structure to the enlarged
Wronskian representation of the general-$n$ problem. We show there
that mixed second derivatives remain within the original enlarged
determinant, whereas pure second derivatives generate an additional
$(2)$ sector requiring a one-step Wronskian extension, whose
external contributions are governed by local three-term Pl\"ucker
relations. Section~7 relates the determinant weight grading to the
zero-order term of the Nakamura operator. Finally, Sec.~8 summarizes
the results and discusses their implications for the Nakamura
conjecture and for a possible trilinear formulation.

\section{Adapted coordinates and hyperbolic structures}
\subsection{Nakamura operator in adapted coordinates}

The differential operator entering the Nakamura conjecture may be
written as
\begin{equation}
 {\cal F}(a\cdot b)
 =
 (x^2-1)D_x^2(a\cdot b)
 +2x\,\partial_x(ab)
 +(y^2-1)D_y^2(a\cdot b)
 +2y\,\partial_y(ab)
 -2n^2ab ,
\label{Fxy}
\end{equation}
where
\begin{equation}
 D_x^2(a\cdot b)
 =
 a_{xx}b-2a_xb_x+ab_{xx}
\end{equation}
and similarly for $D_y^2$.

The Toda determinant representation is generated by the operators Eq.~(\ref{Lpm}).
We now introduce the coordinates defined in Eq.~(\ref{coord}).
They satisfy
\begin{equation}
 (x^2-1)\partial_x=-\partial_\xi,
 \qquad
 (y^2-1)\partial_y=-\partial_\eta .
\label{adaptedder}
\end{equation}
Consequently,
\begin{equation}
       L_+=-\partial_\xi-\partial_\eta,
       \qquad
       L_-=-\partial_\xi+\partial_\eta .
\label{Lpmnew}
\end{equation}
Thus the differential algebra generating the Toda determinants becomes
a constant-coefficient differential algebra.

The same transformation considerably simplifies the Nakamura
operator.  A direct calculation gives
\begin{equation}
 (x^2-1)D_x^2(a\cdot b)
 +2x\,\partial_x(ab)
 =
 \frac{1}{x^2-1}D_\xi^2(a\cdot b),
\end{equation}
and
\begin{equation}
 (y^2-1)D_y^2(a\cdot b)
 +2y\,\partial_y(ab)
 =
 \frac{1}{y^2-1}D_\eta^2(a\cdot b).
\end{equation}
Using
\begin{equation}
 \frac{1}{x^2-1}=\sinh^2\xi,
 \qquad
 \frac{1}{y^2-1}=-\cosh^2\eta ,
\end{equation}
we obtain
\begin{equation}
 \boxed{
 {\cal F}
 =
 \sinh^2\xi\,D_\xi^2
 -
 \cosh^2\eta\,D_\eta^2
 -
 2n^2 .
 }
\label{Fadapted}
\end{equation}

The ordinary first derivatives appearing in Eq.~(\ref{Fxy}) have
therefore disappeared.  The remaining coordinate dependence occurs
only through the multiplicative factors $\sinh^2\xi$ and
$\cosh^2\eta$.

\subsection{Hyperbolic representation of the Toda building blocks}

The Toda determinants are generated by repeated actions of
$L_{+}$ and $L_{-}$ on the elementary seed Eq.~(\ref{psi}).

Since $L_{\pm}$ are linear combinations of the one-variable operators Eq.~(\ref{adaptedder}), the individual matrix elements of the Toda Wronskians are built from
successive applications of these two operators to $x$ and $y$,
respectively.  It is therefore convenient to introduce a one-variable
sequence which keeps track of these repeated differentiations.

Define
\begin{equation}
W_1(z)=z,\qquad
W_{r+1}(z)=(z^2-1)\frac{d}{dz}W_r(z).
\label{Wrec}
\end{equation}
In the adapted coordinates,
\begin{equation}
 W_r(\coth\xi)
 =
 (-\partial_\xi)^{r-1}\coth\xi ,
\label{Wx}
\end{equation}
whereas
\begin{equation}
 W_r(\tanh\eta)
 =
 (-\partial_\eta)^{r-1}\tanh\eta .
\label{Wy}
\end{equation}

Repeated differentiation shows that
\begin{equation}
 W_r(\coth\xi)
 =
 (\sinh\xi)^{-r}
 P_r(e^\xi,e^{-\xi}),
\label{WrX}
\end{equation}
and
\begin{equation}
 W_r(\tanh\eta)
 =
 (\cosh\eta)^{-r}
 Q_r(e^\eta,e^{-\eta}),
\label{WrY}
\end{equation}
where $P_r$ and $Q_r$ are finite Laurent polynomials.
Their explicit forms will not be needed below.  The important point is
that all non-polynomial hyperbolic dependence is isolated in the factors
$(\sinh\xi)^{-r}$ and $(\cosh\eta)^{-r}$, respectively.  This separation
allows us to assign a definite hyperbolic weight $r$ to each Toda
building block.
For example,
\begin{align}
 W_1(\coth\xi)&=\coth\xi,\\
 W_2(\coth\xi)&=\csch^2\xi,\\
 W_3(\coth\xi)&=2\coth\xi\,\csch^2\xi,
\end{align}
while
\begin{align}
 W_1(\tanh\eta)&=\tanh\eta,\\
 W_2(\tanh\eta)&=-\sech^2\eta,\\
 W_3(\tanh\eta)&=-2\tanh\eta\,\sech^2\eta.
\end{align}

This motivates the definition of the hyperbolic weight factor
\begin{equation}
 \rho_{\alpha\beta}(\xi,\eta)
 =
 (\sinh\xi)^{-\alpha}
 (\cosh\eta)^{-\beta}.
\label{rho}
\end{equation}
We define its total weight by
\begin{equation}
       w(\rho_{\alpha\beta})=\alpha+\beta.
\label{weightdef}
\end{equation}

\section{Weight theorem for the Toda determinants}

The Toda tau function is represented by
\begin{equation}
 g_n=\tau_n
 =
 \det\left[
 L_+^{\,i-1}L_-^{\,j-1}\psi
 \right]_{i,j=1}^{n},
\label{gn}
\end{equation}
with elementary seed Eq.~(\ref{psi}).
       In the adapted coordinates,
\begin{equation}
       \psi
       =
       p\coth\xi-iq\tanh\eta .
\label{seedadapted}
\end{equation}

The matrix element in row $i$ and column $j$ involves a function of
order
\begin{equation}
       r_{ij}=i+j-1 .
\end{equation}
According to Eqs.~(\ref{WrX}) and (\ref{WrY}), either its $x$ part
carries weight $r_{ij}$ in $\sinh\xi$, or its $y$ part carries the
same weight in $\cosh\eta$.

Consider an arbitrary term in the permutation expansion of
Eq.~(\ref{gn}).  Expanding the determinant in Eq.~(\ref{gn}) by the Leibniz formula,
each term is specified by a permutation $\sigma\in S_n$ and has
the form
\begin{equation}
 {\rm sgn}(\sigma)\prod_{i=1}^n A_{i,\sigma(i)},
 \qquad
 A_{ij}=L_+^{\,i-1}L_-^{\,j-1}\psi .
\end{equation}
Since the matrix element $A_{i,\sigma(i)}$ has differential order
$i+\sigma(i)-1$, the total weight of this determinant term is
\begin{align}
 w
 &=
 \sum_{i=1}^{n}
 \left(i+\sigma(i)-1\right)
 \nonumber\\
 &=
 \sum_{i=1}^{n}i
 +
 \sum_{i=1}^{n}\sigma(i)
 -n .
\end{align}
Since $\sigma$ is a permutation,
\begin{equation}
 \sum_{i=1}^{n}\sigma(i)
 =
 \sum_{i=1}^{n}i
 =
 \frac{n(n+1)}{2}.
\end{equation}
Therefore
\begin{equation}
 \boxed{
       w(g_n)=n^2 .
 }
\label{weightg}
\end{equation}

More explicitly,
\begin{equation}
 g_n(\xi,\eta)
 =
 \sum_{\alpha+\beta=n^2}
 \rho_{\alpha\beta}
 P_{\alpha\beta}^{(n)}
 (e^{\pm\xi},e^{\pm\eta}),
\label{gdecomp}
\end{equation}
where each $P_{\alpha\beta}^{(n)}$ is a finite Laurent polynomial.

The second Toda function may be written as
\begin{equation}
 f_n
 =
 \tau_{n-1}
 \big|_{\psi\rightarrow L_+L_-\psi}.
\label{fn}
\end{equation}
It is therefore an $(n-1)\times(n-1)$ determinant whose matrix
elements have order
\begin{equation}
       r_{ij}=i+j+1.
\end{equation}
Putting $m=n-1$, the total weight of any determinant term is
\begin{align}
 w(f_n)
 &=
 \sum_{i=1}^{m}
 [i+\sigma(i)+1]
 \nonumber\\
 &=
 m(m+1)+m
 \nonumber\\
 &=
 m(m+2)
 \nonumber\\
 &=
 \boxed{n^2-1}.
\label{weightf}
\end{align}

We have therefore obtained the general weight relations
\begin{equation}
 \boxed{
 w(g_n)=n^2,\qquad
 w(f_n)=n^2-1.
 }
\label{weighttheorem}
\end{equation}

In particular,
\begin{equation}
 w(g_n)-w(f_n)=1.
 \label{unitweight}
\end{equation}
The ratio $g_n/f_n$, which enters the Ernst potential, thus carries a
natural relative weight one.
The significance of this weight goes beyond the counting of
hyperbolic factors.  Graded structures occur naturally in soliton
theory and, in the Sato formulation of integrable hierarchies, the
tau functions are organized by partitions and the associated
Pluecker coordinates \cite{Sato1981, MiwaJimboDate}.  The degree of a partition is measured by
the number of boxes, and the addition of one box raises this degree
by one.  The hyperbolic weight introduced here is not assumed
a priori to be identical to this standard grading.  Nevertheless,
the Toda determinants considered here exhibit precisely the same
incremental structure: one action of a generating operator raises
the hyperbolic weight by one and, as will be shown in Sec.~5,
corresponds to the addition of one box to the associated partition.

Thus the weight found in Eqs.~(\ref{weighttheorem}) and (\ref{unitweight}) should not be regarded
merely as a bookkeeping device.  It provides a grading of the Toda
determinant hierarchy which is compatible with its partition and
Pluecker structures.  A further indication of its significance will
appear in Sec.~7, where the same integer $n^2$ is shown to be directly
related to the zero-order term of the Nakamura bilinear operator.

\section{Examples of the weight structure}
\subsection{Nonrotating sector}
For $q=0$ the $y$ dependence disappears and only the sector
\begin{equation}
       (\alpha,\beta)=(n^2,0)
\end{equation}
survives.

The two elementary functions appearing in the closed form of the
nonrotating Toda solution can be written as
\begin{equation}
 u_n
 =
 (x^2-1)^{n(n-1)/2}(x+1)^n,
\end{equation}
and
\begin{equation}
 v_n
 =
 (x^2-1)^{n(n-1)/2}(x-1)^n .
\end{equation}
Using $x=\coth\xi$ gives
\begin{equation}
 \boxed{
 u_n
 =
 \frac{e^{n\xi}}{\sinh^{n^2}\xi},
 \qquad
 v_n
 =
 \frac{e^{-n\xi}}{\sinh^{n^2}\xi}.
 }
\label{uv}
\end{equation}

Thus the nonrotating solution consists of two ordinary exponentials
$e^{\pm n\xi}$ propagating on a common hyperbolic background
$\sinh^{-n^2}\xi$.

This observation explains the simple logarithmic derivative relations
used in the conventional proof:
\begin{equation}
 \partial_\xi\log u_n
 =
 n-n^2\coth\xi,
\end{equation}
\begin{equation}
 \partial_\xi\log v_n
 =
 -n-n^2\coth\xi.
\end{equation}
The difference of the two logarithmic derivatives is constant,
\begin{equation}
 \partial_\xi\log u_n
 -
 \partial_\xi\log v_n
 =
 2n.
\end{equation}
This is precisely the type of relation for which the $Z_3$-weighted
trilinear operators acquire a simple spectral interpretation.

\subsection{Rotating $n=2$ solution}

For $n=2$ the Toda functions can be written as
\begin{equation}
 g_2
 =
 p^2(x^4-1)
 +q^2(y^4-1)
 +2ipq\,xy(y^2-x^2),
\label{g2xy}
\end{equation}
and
\begin{equation}
 f_2
 =
 2\left[
 px(x^2-1)
 +iqy(y^2-1)
 \right].
\label{f2xy}
\end{equation}

In the adapted coordinates,
\begin{align}
 g_2={}&
 p^2\rho_{40}\cosh2\xi
 -q^2\rho_{04}\cosh2\eta
 \nonumber\\
 &-2ipq\,\rho_{13}\cosh\xi\sinh\eta
 -2ipq\,\rho_{31}\cosh\xi\sinh\eta .
\label{g2weight}
\end{align}
Every term satisfies
\begin{equation}
       \alpha+\beta=4=n^2 .
\end{equation}

The second tau function is
\begin{equation}
 f_2
 =
 2p\,\rho_{30}\cosh\xi
 -
 2iq\,\rho_{03}\sinh\eta ,
\label{f2weight}
\end{equation}
and every term satisfies
\begin{equation}
       \alpha+\beta=3=n^2-1.
\end{equation}

The $n=2$ solution therefore provides an explicit realization of the
general weight theorem (\ref{weighttheorem}).

The numerators appearing in Eq.~(\ref{g2weight}) are finite
exponential sums.  For example,
\begin{equation}
 \cosh\xi\sinh\eta
 =
 \frac14
 \left(
 e^{\xi+\eta}
 -e^{\xi-\eta}
 +e^{-\xi+\eta}
 -e^{-\xi-\eta}
 \right).
\end{equation}
The corresponding spectral points include
\begin{equation}
 (\pm2,0),\qquad
 (0,\pm2),\qquad
 (\pm1,\pm1).
\end{equation}
Thus the rotating solution may be viewed as a finite set of
exponential modes distributed among different hyperbolic weight
sectors.

\section{Toda determinant lattice and shifted minors}
\label{sec:toda_lattice}

The weight theorem derived in Sec.~3 can be understood more naturally
by embedding the two functions $g_n$ and $f_n$ into a larger family of
shifted Toda determinants.  This viewpoint is useful because the
Nakamura conjecture asserts that the TS functions are
generated by Toda-molecule determinants for arbitrary $n$.  It is
therefore natural to regard the determinant hierarchy itself, rather
than a particular differential ansatz, as the primary organizing
principle.

Let
\begin{equation}
 a_{ij}=L_+^i L_-^j\psi ,
 \qquad i,j=0,1,2,\ldots ,
 \label{eq:aij}
\end{equation}
where
\begin{equation}
 \psi=p\coth\xi-iq\tanh\eta .
 \label{eq:seed7}
\end{equation}
Since the adapted coordinates introduced in Sec.~2 give
\begin{equation}
 L_+=-\partial_\xi-\partial_\eta,
 \qquad
 L_-=-\partial_\xi+\partial_\eta ,
 \label{eq:Ladapted7}
\end{equation}
the two operators commute,
\begin{equation}
 [L_+,L_-]=0 .
 \label{eq:Lcommute}
\end{equation}

It is useful first to separate the two shift directions.  Define
\begin{equation}
 \boxed{
 \tau_m^{[r,s]}
 =
 \det\left[
 L_+^{\,i+r}L_-^{\,j+s}\psi
 \right]_{i,j=0}^{m-1}
 }
 \qquad (r,s\geq0).
 \label{eq:twoshifttau}
\end{equation}
The common-shift determinants used below are the diagonal subset
\begin{equation}
 \tau_m^{[r]}=\tau_m^{[r,r]}.
 \label{eq:commonshift}
\end{equation}
For later use we refine Eq.~(\ref{eq:twoshifttau}) by partitions
$\lambda=(\lambda_1\geq\cdots\geq\lambda_m\geq0)$ and
$\mu=(\mu_1\geq\cdots\geq\mu_m\geq0)$:
\begin{equation}
 \boxed{
 \tau_{m;\lambda,\mu}^{[r,s]}
 =
 \det\!\left[
 L_+^{\,r+i+\lambda_{m-i}}
 L_-^{\,s+j+\mu_{m-j}}\psi
 \right]_{i,j=0}^{m-1} .
 }
 \label{eq:partitiontau}
\end{equation}
The empty partitions reproduce Eq.~(\ref{eq:twoshifttau}).  This
notation records the nonuniform row and column shifts which are
created by differentiation and which cannot in general be represented
by a uniform shift of all rows or columns.

The Wronskian structure gives an exact derivative rule.  Differentiating
with $L_+$ replaces one row index by the next one.  A term vanishes
whenever the shifted row coincides with its neighbor; the surviving
terms are exactly the addable boxes of the partition $\lambda$.
Hence
\begin{equation}
 \boxed{
 L_+\tau_{m;\lambda,\mu}^{[r,s]}
 =
 \sum_{\lambda+\square}
 \tau_{m;\lambda+\square,\mu}^{[r,s]} ,
 }
 \label{eq:boxplus}
\end{equation}
where the sum runs over all Young diagrams obtained from $(\lambda)$ by adding one admissible box, and similarly
\begin{equation}
 \boxed{
 L_-\tau_{m;\lambda,\mu}^{[r,s]}
 =
 \sum_{\mu+\square}
 \tau_{m;\lambda,\mu+\square}^{[r,s]} .
 }
 \label{eq:boxminus}
\end{equation}
To see this explicitly, differentiate the determinant row by row.
For a given row, the action of $L_+$ increases its row index by one.
If the resulting index coincides with that of the adjacent row, the
corresponding determinant vanishes by antisymmetry.  The nonvanishing
terms are therefore in one-to-one correspondence with the admissible
additions of a box to $\lambda$.  This proves Eq.~(\ref{eq:boxplus}); Eq.~(\ref{eq:boxminus})
follows identically by differentiating column by column. This row--column box-addition rule is illustrated schematically
in Fig.~\ref{fig:young}. For the empty partition this gives, in particular,
\begin{align}
 L_+^2\tau_{m;\varnothing,\varnothing}^{[r,s]}
 &=
 \tau_{m;(1,1),\varnothing}^{[r,s]}
 +
 \tau_{m;(2),\varnothing}^{[r,s]},
 \label{eq:secondpluspartition}
 \\
 L_+L_-\tau_{m;\varnothing,\varnothing}^{[r,s]}
 &=
 \tau_{m;(1),(1)}^{[r,s]},
 \label{eq:mixedpartition}
 \\
 L_-^2\tau_{m;\varnothing,\varnothing}^{[r,s]}
 &=
 \tau_{m;\varnothing,(1,1)}^{[r,s]}
 +
 \tau_{m;\varnothing,(2)}^{[r,s]} .
 \label{eq:secondminuspartition}
\end{align}
For $m=1$ the partition $(1,1)$ is absent, as it should be.

The same notation refines the weight theorem.  Every determinant term
in Eq.~(\ref{eq:partitiontau}) has the same total differential order,
which gives
\begin{equation}
 \boxed{
 w\!\left(\tau_{m;\lambda,\mu}^{[r,s]}\right)
 =m(1+r+s)+2\sum_{i=0}^{m-1}i+|\lambda|+|\mu|
 =m(m+r+s)+|\lambda|+|\mu| .
 }
 \label{eq:partitionweight}
\end{equation}
Thus each action of $L_+$ or $L_-$ adds one box and raises the weight
by one.  The grading and the derivative lattice are therefore two
aspects of the same determinant structure.

We now restrict temporarily to the common-shift subfamily defined in Eq.~(\ref{eq:twoshifttau}).
The ordinary Toda molecule corresponds to the unshifted line
$r=0$,
\begin{equation}
 \tau_m^{[0]}
 =
 \det\left[
 L_+^{\,i}L_-^{\,j}\psi
 \right]_{i,j=0}^{m-1}.
 \label{eq:tau0}
\end{equation}
In particular,
\begin{equation}
 \boxed{
 g_n=\tau_n^{[0]} .
 }
 \label{eq:gshifted}
\end{equation}

Using Eq.~(\ref{fntau}) and the commutativity of \(L_+\) and \(L_-\), one finds
\begin{equation}
 f_n
 =
 \det\left[
 L_+^{\,i+1}L_-^{\,j+1}\psi
 \right]_{i,j=0}^{n-2},
 \label{eq:fshifted1}
\end{equation}
and hence
\begin{equation}
 \boxed{
 f_n=\tau_{n-1}^{[1]} .
 }
 \label{eq:fshifted}
\end{equation}
Thus $g_n$ and $f_n$ are not unrelated determinants.  They occupy
neighboring positions in a two-dimensional lattice labelled by the
determinant size $m$ and the common shift $r$:
\begin{equation}
 g_n:(m,r)=(n,0),
 \qquad
 f_n:(m,r)=(n-1,1).
 \label{eq:latticepoints}
\end{equation}

This representation also gives a simple generalization of the weight
theorem.  The matrix element in row $i$ and column $j$ of
$\tau_m^{[r]}$ has differential order
\begin{equation}
 \nu_{ij}=i+j+2r+1 .
 \label{eq:nuij}
\end{equation}
Here the final $+1$ is the weight of the elementary seed itself.
According to the hyperbolic representation derived in Sec.~2, each
such matrix element is a finite Laurent polynomial multiplied by a
hyperbolic factor of total weight $\nu_{ij}$.

Consider now a term in the determinant expansion corresponding to a
permutation $\sigma\in S_m$.  Its total weight is
\begin{align}
 w\!\left(\tau_m^{[r]}\right)
 &=
 \sum_{i=0}^{m-1}
 \left(
 i+\sigma(i)+2r+1
 \right)
 \nonumber\\
 &=
 \sum_{i=0}^{m-1}i
 +
 \sum_{i=0}^{m-1}\sigma(i)
 +
 m(2r+1).
 \label{eq:weightcalc}
\end{align}
Since $\sigma$ is a permutation,
\begin{equation}
 \sum_{i=0}^{m-1}\sigma(i)
 =
 \sum_{i=0}^{m-1}i
 =
 \frac{m(m-1)}{2},
 \label{eq:permsum7}
\end{equation}
and therefore
\begin{equation}
 \boxed{
 w\!\left(\tau_m^{[r]}\right)
 =
 m(m+2r).
 }
 \label{eq:generalweight}
\end{equation}

The weight relations obtained in Sec.~3 are immediate consequences
of Eq.~(\ref{eq:generalweight}).  Indeed, substituting $(m,r)=(n,0)$ for
$g_n=\tau_n^{[0]}$ gives $w(g_n)=n^2$, while substituting
$(m,r)=(n-1,1)$ for $f_n=\tau_{n-1}^{[1]}$ gives
$w(f_n)=(n-1)(n+1)=n^2-1$.  Thus the unit weight difference
found in Eq.~(\ref{unitweight}) is not an isolated numerical observation, but
follows directly from the relative positions of $g_n$ and $f_n$
in the shifted Toda determinant lattice.

For a fixed TS index $n$, the Nakamura pair is the
diagonal pair
\begin{equation}
 \boxed{
 \tau_n^{[0]}
 \longleftrightarrow
 \tau_{n-1}^{[1]} .
 }
 \label{eq:diagonalpair}
\end{equation}
This diagonal relation is distinct from the usual Toda-molecule
evolution in the determinant-size direction.

Indeed, for a Wronskian-type Toda determinant, Jacobi's determinant
identity gives the familiar Toda relation
\begin{equation}
 \tau_m^{[r]}
 L_+L_-\tau_m^{[r]}
 -
 \left(L_+\tau_m^{[r]}\right)
 \left(L_-\tau_m^{[r]}\right)
 =
 \tau_{m+1}^{[r]}\tau_{m-1}^{[r]},
 \label{eq:todajacobi}
\end{equation}
up to the normalization convention used for the Toda variables.
Equation~(\ref{eq:todajacobi}) connects neighboring determinants in
the horizontal $m$ direction, whereas the Nakamura construction
relates determinants lying on different $r$ levels.

This observation suggests that the unresolved rotating part of the
Nakamura conjecture should be viewed as a problem involving relations
between different directions in the determinant lattice. In particular, the Nakamura pair in Eqs.~(\ref{eq:gshifted}) and (\ref{eq:fshifted})
should be treated together with the adjacent minors generated by
$L_+$ and $L_-$.  Equations~(\ref{eq:boxplus}) and
(\ref{eq:boxminus}) show that repeated derivatives move on the
partition-resolved row/column lattice rather than merely on the
common-shift sublattice.

This viewpoint changes the role of the trilinear formulation.  Rather
than postulating a three-slot differential equation and asking whether
the Toda functions satisfy it, one may start from the Toda determinant
hierarchy and determine which algebraic identities among its minors
reproduce the Nakamura equations.  The relevant identities are
expected to be of Jacobi, Pluecker, or Pfaffian type.

The unit weight relation
\begin{equation}
 n^2=(n^2-1)+1
 \label{eq:unitweight7}
\end{equation}
still makes the triple
\begin{equation}
 (g_n^*,f_n,\psi)
 \label{eq:triple7}
\end{equation}
a natural weight-balanced object.  However, the determinant-lattice
picture shows that weight balance alone does not determine a closed
three-slot differential equation.  A more fundamental possibility is
that any genuine $Z_3$ structure is realized at the level of
relations among Toda determinants and their minors.

In the next section we therefore return to the enlarged Wronskian
representation introduced in the earlier proof of the first
Nakamura equations and examine how the quantities entering the
remaining equations can be organized as minors of a common
determinant.
\section{Enlarged Wronskian and the general-$n$ problem}
\label{sec:enlarged}

The determinant-lattice picture of the preceding section suggests
that the unresolved part of the Nakamura conjecture should be studied
directly at the level of determinant minors.  This viewpoint is also
consistent with the earlier proof of the first pair of Nakamura
equations, where $g_n$, $f_n$ and their complex conjugates were
embedded into a common enlarged Wronskian determinant.

Following Ref.~\cite{FKY}, let $D$ denote the $(n+2)\times(n+2)$
enlarged determinant constructed from the Toda seed $\psi$, its
complex conjugate, and successive actions of the Toda generating
operators.  We use the notation
\begin{equation}
 D\begin{bmatrix}
 i_1,\ldots,i_r\\
 j_1,\ldots,j_r
 \end{bmatrix}
\label{minor-notation}
\end{equation}
for the minor obtained by deleting rows $i_1,\ldots,i_r$ and columns
$j_1,\ldots,j_r$.

In this notation the four functions entering the Nakamura equations
are represented as
\begin{equation}
 g_n=
 D\begin{bmatrix}
 1,n+2\\
 n+1,n+2
 \end{bmatrix},
 \qquad
 f_n=
 D\begin{bmatrix}
 1,2,n+2\\
 1,n+1,n+2
 \end{bmatrix},
\label{gf-minors}
\end{equation}
and
\begin{equation}
 g_n^*=
 D\begin{bmatrix}
 n+1,n+2\\
 1,n+2
 \end{bmatrix},
 \qquad
 f_n^*=
 D\begin{bmatrix}
 1,n+1,n+2\\
 1,2,n+2
 \end{bmatrix}.
\label{gfstar-minors}
\end{equation}
Thus the complex-conjugate functions need not be treated as
determinants of a separate matrix: all four functions are minors of
the same enlarged determinant $D$.

The Wronskian structure also gives the first derivatives as minors of
the same determinant.  In particular,
\begin{equation}
 L_+g_n=
 D\begin{bmatrix}
 1,n+1\\
 n+1,n+2
 \end{bmatrix},
 \qquad
 L_-g_n=
 D\begin{bmatrix}
 1,n+2\\
 n,n+2
 \end{bmatrix},
\label{g-first}
\end{equation}
\begin{equation}
 L_+f_n=
 D\begin{bmatrix}
 1,2,n+1\\
 1,n+1,n+2
 \end{bmatrix},
 \qquad
 L_-f_n=
 D\begin{bmatrix}
 1,2,n+2\\
 1,n,n+2
 \end{bmatrix},
\label{f-first}
\end{equation}
and
\begin{equation}
 L_+g_n^*=
 D\begin{bmatrix}
 n,n+2\\
 1,n+2
 \end{bmatrix},
 \qquad
 L_-g_n^*=
 D\begin{bmatrix}
 n+1,n+2\\
 1,n+1
 \end{bmatrix}.
\label{gstar-first}
\end{equation}
These relations were the starting point of the Pfaffian proof of the
first pair of Nakamura equations in Ref.~\cite{FKY}.

\subsection{Mixed second derivatives}

The remaining equations contain second derivatives.  It is useful to
separate mixed and pure second derivatives.  Let
\begin{equation}
 G=g_n^*,\qquad H=f_n.
\label{GHdef}
\end{equation}
The mixed Hirota derivative is
\begin{equation}
 D_+D_-(G\cdot H)
 =
 G_{+-}H-G_+H_- -G_-H_+ +GH_{+-},
\label{mixed-Hirota}
\end{equation}
where a subscript $+$ or $-$ denotes the action of $L_+$ or $L_-$.

Because $L_+$ and $L_-$ generate row and column shifts of the
Wronskian matrix, the mixed derivatives remain within the same
$(n+2)\times(n+2)$ enlarged determinant.  In particular,
\begin{equation}
 G_{+-}=
 D\begin{bmatrix}
 n,n+2\\
 1,n+1
 \end{bmatrix},
 \qquad
 H_{+-}=
 D\begin{bmatrix}
 1,2,n+1\\
 1,n,n+2
 \end{bmatrix}.
\label{mixed-second}
\end{equation}
Consequently,
\begin{align}
 {\cal M}_n
 &\equiv D_+D_-(g_n^*\cdot f_n)
\nonumber\\
 &=
 D\begin{bmatrix}
 n,n+2\\
 1,n+1
 \end{bmatrix}
 D\begin{bmatrix}
 1,2,n+2\\
 1,n+1,n+2
 \end{bmatrix}
\nonumber\\
 &\quad-
 D\begin{bmatrix}
 n,n+2\\
 1,n+2
 \end{bmatrix}
 D\begin{bmatrix}
 1,2,n+2\\
 1,n,n+2
 \end{bmatrix}
\nonumber\\
 &\quad-
 D\begin{bmatrix}
 n+1,n+2\\
 1,n+1
 \end{bmatrix}
 D\begin{bmatrix}
 1,2,n+1\\
 1,n+1,n+2
 \end{bmatrix}
\nonumber\\
 &\quad+
 D\begin{bmatrix}
 n+1,n+2\\
 1,n+2
 \end{bmatrix}
 D\begin{bmatrix}
 1,2,n+1\\
 1,n,n+2
 \end{bmatrix}.
\label{mixed-M}
\end{align}

Equation~(\ref{mixed-M}) is important because the mixed second-order
part is already closed within the enlarged determinant used in the
earlier first-order proof.  No new tau function is required.

The structure of Eq.~(\ref{mixed-M}) is also reminiscent of the
minor identities used in Ref.~\cite{FKY}.  A useful three-term
identity is
\begin{equation}
 A\begin{bmatrix}a,b\\d,e\end{bmatrix}
 A\begin{bmatrix}c\\e\end{bmatrix}
 +
 A\begin{bmatrix}b,c\\d,e\end{bmatrix}
 A\begin{bmatrix}a\\e\end{bmatrix}
 +
 A\begin{bmatrix}c,a\\d,e\end{bmatrix}
 A\begin{bmatrix}b\\e\end{bmatrix}
 =0,
\label{pfaffian-identity}
\end{equation}
with the appropriate ordering of the deleted indices.  In the
earlier proof, expressions generated by the Nakamura equations were
reduced by identities of this type, together with a special symmetry
property of the enlarged determinant.

The same procedure can be applied to
Eq.~(\ref{mixed-M}).  The important point for the present discussion
is that its reduction is an entirely algebraic problem involving
minors of a single $(n+2)\times(n+2)$ determinant.

\subsection{Pure second derivatives and one-step enlargement}

The pure second derivatives have a different structure:
\begin{equation}
 D_+^2(G\cdot H)
 =
 G_{++}H-2G_+H_+ +GH_{++},
\label{plus-pure}
\end{equation}
and
\begin{equation}
 D_-^2(G\cdot H)
 =
 G_{--}H-2G_-H_- +GH_{--}.
\label{minus-pure}
\end{equation}
Repeated shifts in the same direction do not, in general, remain
represented by a single minor of $D$.

For compactness, put
\begin{equation}
 N=n+2.
\end{equation}
Direct differentiation of the Wronskian minors gives
\begin{equation}
 G_{++}
 =
 D\begin{bmatrix}
 n-1,N\\
 1,N
 \end{bmatrix}
 +
 D\begin{bmatrix}
 n,n+1\\
 1,N
 \end{bmatrix},
\label{Gpp}
\end{equation}
whereas
\begin{equation}
 H_{++}
 =
 D\begin{bmatrix}
 1,2,n\\
 1,n+1,N
 \end{bmatrix}
 +E_+ .
\label{Hpp}
\end{equation}
Here $E_+$ denotes the contribution obtained when the last surviving
row is shifted once beyond the original Wronskian range.
For example, in differentiating $H_+$ once more with $L_+$,
all shifts producing coincident rows vanish.  Two nonvanishing
possibilities remain: the penultimate surviving row can be shifted,
giving the first minor in Eq.~(\ref{Hpp}), or the last surviving row can
be shifted beyond the range of $D_{n+2}$, giving $E_+$.
Similarly,
\begin{equation}
 G_{--}
 =
 D\begin{bmatrix}
 n+1,N\\
 1,n
 \end{bmatrix}
 +E_- ,
\label{Gmm}
\end{equation}
while
\begin{equation}
 H_{--}
 =
 D\begin{bmatrix}
 1,2,N\\
 1,n-1,N
 \end{bmatrix}
 +
 D\begin{bmatrix}
 1,2,N\\
 1,n,n+1
 \end{bmatrix}.
\label{Hmm}
\end{equation}
Thus the apparent failure of closure of the pure derivatives is
localized in the two terms $E_+$ and $E_-$.

These terms have a simple determinant interpretation.  Let
\begin{equation}
 M=n+3=N+1
\end{equation}
and extend $D$ by one Wronskian row and one Wronskian column to an
$(n+3)\times(n+3)$ determinant $\widehat D$.  The original $D$ is
the upper-left $(n+2)\times(n+2)$ principal block of $\widehat D$.
The same deletion notation will be used for minors of $\widehat D$.

The two external terms are then ordinary minors of $\widehat D$:
\begin{equation}
 E_+
 =
 \widehat D
 \begin{bmatrix}
 1,2,n+1,N\\
 1,n+1,N,M
 \end{bmatrix},
\label{Eplus}
\end{equation}
and
\begin{equation}
 E_-
 =
 \widehat D
 \begin{bmatrix}
 n+1,N,M\\
 1,n+1,N
 \end{bmatrix}.
\label{Eminus}
\end{equation}

The partition notation of Sec.~5 identifies these terms more sharply.
Since
\begin{equation}
 H=f_n=\tau_{n-1;\varnothing,\varnothing}^{[1,1]}[\psi],
 \label{Hpartition}
\end{equation}
Eq.~(\ref{eq:secondpluspartition}) gives
\begin{equation}
 H_{++}
 =
 \tau_{n-1;(1,1),\varnothing}^{[1,1]}[\psi]
 +
 \tau_{n-1;(2),\varnothing}^{[1,1]}[\psi].
 \label{Hpppartition}
\end{equation}
Direct comparison with the surviving row indices in
Eqs.~(\ref{Hpp}) and (\ref{Eplus}) yields
\begin{equation}
 D\begin{bmatrix}
 1,2,n\\
 1,n+1,N
 \end{bmatrix}
 =
 \tau_{n-1;(1,1),\varnothing}^{[1,1]}[\psi],
 \qquad
 \boxed{
 E_+=\tau_{n-1;(2),\varnothing}^{[1,1]}[\psi] } .
 \label{Epluspartition}
\end{equation}
Thus the external $+$ term is exactly the row $(2)$ sector.

Likewise $G=g_n^*$ is a Toda Wronskian with conjugate seed,
\begin{equation}
 G=\tau_{n;\varnothing,\varnothing}^{[0,0]}[\psi^*],
 \label{Gpartition}
\end{equation}
when written in the surviving rows and columns of the FKY enlarged
determinant.  Hence
\begin{equation}
 G_{--}
 =
 \tau_{n;\varnothing,(1,1)}^{[0,0]}[\psi^*]
 +
 \tau_{n;\varnothing,(2)}^{[0,0]}[\psi^*],
 \label{Gmmpartition}
\end{equation}
and comparison with Eqs.~(\ref{Gmm}) and (\ref{Eminus}) gives
\begin{equation}
 D\begin{bmatrix}
 n+1,N\\
 1,n
 \end{bmatrix}
 =
 \tau_{n;\varnothing,(1,1)}^{[0,0]}[\psi^*],
 \qquad
 \boxed{
 E_-=\tau_{n;\varnothing,(2)}^{[0,0]}[\psi^*] } .
 \label{Eminuspartition}
\end{equation}
The need for $\widehat D_{n+3}$ therefore has a simple combinatorial
origin: two shifts in the same direction generate the partition $(2)$,
whose last row or column lies one step outside the range of $D_{n+2}$.
By contrast, the mixed sector $(1),(1)$ remains inside that range.
The functions $G=g_n^*$ and $H=f_n$ themselves may of course be
embedded in the same enlarged determinant:
\begin{equation}
 G=
 \widehat D
 \begin{bmatrix}
 n+1,N,M\\
 1,N,M
 \end{bmatrix},
\qquad
 H=
 \widehat D
 \begin{bmatrix}
 1,2,N,M\\
 1,n+1,N,M
 \end{bmatrix}.
\label{GHhat}
\end{equation}

Equations~(\ref{Eplus})--(\ref{GHhat}) show that the pure
second-order terms do not introduce an unrelated new function.
Rather, they require only a one-step extension of the same Wronskian
hierarchy,
\begin{equation}
 D_{n+2}\ \longrightarrow\ \widehat D_{n+3}.
\label{one-step}
\end{equation}

The Pfaffian identity (\ref{pfaffian-identity}) can now also be
applied to the external products.  Define
\begin{equation}
 K_1=
 D\begin{bmatrix}
 2,n+1,N\\
 1,n+1,N
 \end{bmatrix},
\qquad
 K_2=
 D\begin{bmatrix}
 1,n+1,N\\
 1,n+1,N
 \end{bmatrix},
\label{K12}
\end{equation}
and
\begin{align}
 R_1^{(+)}
 &=
 \widehat D
 \begin{bmatrix}
 1,n+1,N\\
 1,N,M
 \end{bmatrix},
&
 R_1^{(-)}
 &=
 \widehat D
 \begin{bmatrix}
 1,N,M\\
 1,n+1,N
 \end{bmatrix},
\label{R1}
\\
 R_2^{(+)}
 &=
 \widehat D
 \begin{bmatrix}
 2,n+1,N\\
 1,N,M
 \end{bmatrix},
&
 R_2^{(-)}
 &=
 \widehat D
 \begin{bmatrix}
 2,N,M\\
 1,n+1,N
 \end{bmatrix}.
\label{R2}
\end{align}
Application of Eq.~(\ref{pfaffian-identity}) gives
\begin{equation}
 GE_+
 =
 -K_1R_1^{(+)}
 -K_2R_2^{(+)},
\label{GEplus}
\end{equation}
and
\begin{equation}
 E_-H
 =
 -K_1R_1^{(-)}
 -K_2R_2^{(-)}.
\label{EminusH}
\end{equation}

These two reductions are instances of the same local Pluecker cell.
For a fixed background set of deleted rows and columns, denote by
$\Delta_a$ a maximal minor obtained by deleting one of three variable
rows $a,b,c$, and by $\delta_{bc}$ the corresponding minor obtained
by deleting the complementary pair.  With the ordered-minor convention
of Eq.~(\ref{pfaffian-identity}), the local relation is
\begin{equation}
 \boxed{
 \Delta_a\,\delta_{bc}
 +\Delta_b\,\delta_{ca}
 +\Delta_c\,\delta_{ab}=0 .
 }
 \label{local-pluecker}
\end{equation}
For Eq.~(\ref{GEplus}) the three variable row labels may be chosen as
\begin{equation}
 (a,b,c)=(M,1,2),
 \label{pluscell}
\end{equation}
so that
\begin{equation}
 \Delta_M=G,\qquad
 \Delta_1=R_1^{(+)},\qquad
 \Delta_2=R_2^{(+)},
\end{equation}
and
\begin{equation}
 \delta_{12}=E_+,\qquad
 \delta_{2M}=K_1,\qquad
 \delta_{M1}=K_2,
\end{equation}
with signs fixed by the same ordering convention.  Equation
(\ref{local-pluecker}) then reproduces Eq.~(\ref{GEplus}).
For Eq.~(\ref{EminusH}) the identical local relation is obtained after
interchanging the row/column role appropriate to the $-$ sector.
Thus the two external reductions are not independent coincidences:
they are the two orientations of the same three-term Pluecker geometry.
Consequently, the complete contribution containing the external
Wronskian shifts reduces to
\begin{equation}
 GE_+ +E_-H
 =
 -K_1S_1-K_2S_2,
\label{external-reduction}
\end{equation}
where
\begin{equation}
 S_1=R_1^{(+)}+R_1^{(-)},
 \qquad
 S_2=R_2^{(+)}+R_2^{(-)}.
\label{S12}
\end{equation}

This reduction is useful because it localizes the remaining
one-step extension in two definite combinations of minors.  We do
not assume a general row--column transpose symmetry for these
minors.  The symmetry relation used in the earlier proof of
Ref.~\cite{FKY} applies to a special deletion pattern and does
not by itself imply that either $S_1$ or $S_2$ vanishes.

\subsection{Reduction of the Nakamura equation}

We finally return to the adapted coordinates introduced in Sec.~2.
Since
\begin{equation}
 D_\xi=-\frac{1}{2}(D_++D_-),
 \qquad
 D_\eta=-\frac{1}{2}(D_+-D_-),
\label{DxiDeta}
\end{equation}
we have
\begin{equation}
 D_\xi^2
 =
 \frac{1}{4}
 \left(D_+^2+2D_+D_-+D_-^2\right),
\label{Dxisq}
\end{equation}
and
\begin{equation}
 D_\eta^2
 =
 \frac{1}{4}
 \left(D_+^2-2D_+D_-+D_-^2\right).
\label{Detasq}
\end{equation}
The Nakamura operator therefore becomes
\begin{align}
 {\cal F}
 ={}&
 \frac{1}{4}
 \left(\sinh^2\xi-\cosh^2\eta\right)
 \left(D_+^2+D_-^2\right)
\nonumber\\
 &+
 \frac{1}{2}
 \left(\sinh^2\xi+\cosh^2\eta\right)
 D_+D_-
 -2n^2 .
\label{Fpm}
\end{align}

Equation~(\ref{Fpm}) makes the determinant structure of the
general-$n$ problem explicit.  The mixed part $D_+D_-$ is closed
within the original $(n+2)\times(n+2)$ enlarged determinant $D$.
The pure part $D_+^2+D_-^2$ closes after the one-step Wronskian
extension $D_{n+2}\rightarrow\widehat D_{n+3}$, with its external
contribution reduced to the two combinations $S_1$ and $S_2$ in
Eq.~(\ref{S12}).

Thus the remaining rotating problem has been reduced to a definite
minor-identity problem in the one-step enlarged Wronskian hierarchy.
A complete proof would require showing that these minor combinations,
together with the terms already contained in $D$, combine with the
coordinate factors in Eq.~(\ref{Fpm}) and the zero-order term
$-2n^2$ to give the required Nakamura equation.  No such cancellation
is assumed here. The unresolved problem is therefore no longer the unrestricted
differentiation of the Toda determinants.  It is reduced to checking
the cancellation of the finite set of minors generated by the
$(1,1)$, $(2)$, and mixed sectors appearing in Eq.~(\ref{Fpm}). An instructive consequence of this reduction is obtained by considering
the first few values of the TS index.  For $n=2$ one has
$f_2=\tau^{[1]}_1$, so that the $(1,1)$ partition is absent and the
second-order derivative structure is degenerate.  The case $n=3$ is
therefore the first nontrivial member for which all the sectors
\[
 \emptyset,\qquad (1,1),\qquad (2),\qquad ((1),(1))
\]
occur simultaneously.  In this sense the $n=3$ solution represents the
first complete local cell of the partition-resolved determinant
hierarchy.  Since the derivative rules generating these sectors are
independent of the determinant size, the same local structure persists
for arbitrary $n$.  This observation also gives a structural interpretation of the
finite-$n$ verification by Nakamura and Ohta.  Their Pfaffian
expressions were explicitly checked through $n=6$, with the
$n=5$ and $6$ cases verified by REDUCE3 \cite{NakamuraOhta}.
From the present viewpoint, these cases are not associated with
successively new second-order derivative sectors: for $n\geq3$
they realize the same local partition structure at increasing
determinant size.  The extension from the verified finite cases
to arbitrary $n$ is therefore reduced to establishing the
corresponding universal minor identity. The remaining general-$n$ problem is consequently
not the generation of further derivative sectors, but the identification
of the Pluecker combination which incorporates these sectors together
with the coordinate coefficients and the zeroth-order term in the
Nakamura operator.

The adapted coordinates and the determinant grading play
complementary roles in this formulation.  The transformation
\[
 x=\coth\xi,\qquad y=\tanh\eta
\]
turns $L_\pm$ into constant-coefficient row and column shift
operators, while the remaining coordinate dependence of the
Nakamura operator is carried by the multiplicative factors
$\sinh^2\xi$ and $\cosh^2\eta$.  The partition rule of Sec.~5 tells
which minors are generated by each derivative, the weight formula
constrains their grading, and Eq.~(\ref{local-pluecker}) supplies the
local algebraic reduction.  In this sense the adapted-coordinate,
weight-graded, Young-lattice, and Pfaffian/Pluecker structures are
four aspects of the same Toda determinant hierarchy.
\section{Weight grading and the Nakamura operator}
\label{sec:background}

The relation between the weight grading and the constant term
$-2n^2$ in the Nakamura operator can be made more explicit.

Consider
\begin{equation}
 \rho_{\alpha\beta}
 =
 (\sinh\xi)^{-\alpha}
 (\cosh\eta)^{-\beta}
 =
 e^{\theta(\xi,\eta)}.
\end{equation}
Then
\begin{equation}
 \theta_{\xi\xi}
 =
 \alpha\,\csch^2\xi,
\qquad
 \theta_{\eta\eta}
 =
 -\beta\,\sech^2\eta.
\end{equation}
Consequently,
\begin{equation}
 \sinh^2\xi\,\theta_{\xi\xi}
 -
 \cosh^2\eta\,\theta_{\eta\eta}
 =
 \alpha+\beta.
\label{backgroundweight}
\end{equation}

For a common background multiplying both slots of a bilinear
expression one has
\begin{equation}
 D_\xi^2(\rho A\cdot\rho B)
 =
 \rho^2
 \left[
 D_\xi^2(A\cdot B)
 +2\theta_{\xi\xi}AB
 \right],
\end{equation}
and similarly in the $\eta$ direction.  Hence
\begin{align}
 {\cal F}(\rho A\cdot\rho B)
 =
 \rho^2\bigl\{
 &\sinh^2\xi D_\xi^2(A\cdot B)
 -\cosh^2\eta D_\eta^2(A\cdot B)
 \nonumber\\
 &+
 2(\alpha+\beta-n^2)AB
 \bigr\}.
\label{gaugetransform}
\end{align}
For the common hyperbolic background to be compatible with the
Nakamura bilinear equation without generating an additional
zeroth-order contribution, the last term in Eq.~(\ref{gaugetransform}) must vanish.
This requires
\begin{equation}
 \alpha+\beta=n^2 .
 \label{n2}
\end{equation}

In the nonrotating sector the natural background of $g_n$ is
precisely $\sinh^{-n^2}\xi$, and condition (\ref{n2}) is automatically
satisfied. In the rotating system, however, the determinant weight
is distributed among different $(\alpha,\beta)$ sectors, while
$g_n$ and $f_n$ carry the distinct total weights $n^2$ and
$n^2-1$, respectively.

The appearance of the same integer $n^2$ both in the determinant
grading and in the zero-order term of the Nakamura operator is
therefore not accidental. It indicates that the differential
operator and the Toda determinant representation possess compatible
gradings. This compatibility provides an additional constraint on
the determinant-minor identities required for a general-$n$ proof
of the Nakamura conjecture.
Together with $w(f_n)=n^2-1$ and the unit weight
$w(\psi)=1$ of the elementary Toda seed, this gives
\begin{equation}
 w(g_n)=w(f_n)+w(\psi).
\end{equation}
Thus the same grading that matches the zeroth-order term of the
Nakamura operator also underlies the weight-balanced triple
$(g_n^*,f_n,\psi)$ identified in Eqs.~(\ref{eq:unitweight7}) and (\ref{eq:triple7}).
This observation also complements the determinant analysis of
Sec.~6.  In the adapted coordinates, the operators $L_\pm$ generate
constant-coefficient row and column shifts, whereas the remaining
coordinate dependence of the Nakamura operator is carried by the
multiplicative factors $\sinh^2\xi$ and $\cosh^2\eta$.  Thus a
general determinant proof must respect simultaneously the minor
relations generated by $L_\pm$ and the hyperbolic weight grading.
\section{Discussion and conclusions}
\label{sec:discussion}

We have investigated the Nakamura Toda system from the combined
viewpoints of adapted coordinates, weight grading, and the determinant
hierarchy.  The coordinate transformation
%\begin{equation}
\[
x=\coth\xi,\qquad y=\tanh\eta
%\end{equation}
\]
turns the Toda generating operators into the constant-coefficient
derivatives
%\begin{equation}
 \[
 L_+=-\partial_\xi-\partial_\eta,\qquad
 L_-=-\partial_\xi+\partial_\eta,
%\end{equation}
\]

while the Nakamura operator assumes the form
%\begin{equation}
\[ {\cal F}=\sinh^2\xi\,D_\xi^2-\cosh^2\eta\,D_\eta^2-2n^2.
\]
Thus the ordinary first-derivative terms disappear and the remaining
coordinate dependence is carried only by the multiplicative
hyperbolic factors.

The Toda building blocks are correspondingly represented by
hyperbolic background factors multiplied by finite Laurent
polynomials.  From the determinant representation we obtained the
exact weight relations
%\begin{equation}
\[
 w(g_n)=n^2,\qquad w(f_n)=n^2-1,
%\end{equation}
\]
and, more generally,
%\begin{equation}
\[
 w\!\left(\tau_m^{[r]}\right)=m(m+2r).
%\end{equation}
\]
The relation
%\begin{equation}
 \[
 g_n=\tau_n^{[0]},\qquad
 f_n=\tau_{n-1}^{[1]}
%\end{equation}
\]
shows that the two functions entering the Ernst potential occupy
neighboring diagonal positions in the shifted Toda determinant
lattice.  The rotating $n=2$ solution gives an explicit realization
of this grading, while in the nonrotating sector the full weight
$n^2$ is concentrated in the $\xi$ direction.

The same grading is reflected in the Nakamura differential operator.
As shown in Sec.~7, a common hyperbolic background of total weight
$n^2$ cancels the zero-order term $-2n^2$.  The occurrence of the
same integer $n^2$ in the determinant grading and in the differential
operator therefore expresses a compatibility between the Toda
determinant hierarchy and the Nakamura equation.

The enlarged-Wronskian analysis provides a more direct connection
with the unresolved part of the Nakamura conjecture.  In the
construction of Ref.~\cite{FKY}, the functions $g_n$, $f_n$, $g_n^*$,
and $f_n^*$ and their first derivatives are represented as minors of
a common $(n+2)\times(n+2)$ determinant $D$.  The present analysis
shows that the second derivatives are organized by the partition
lattice introduced in Sec.~5.

The mixed derivative corresponds to the sector $((1),(1))$ and remains
inside $D_{n+2}$.  A pure second derivative splits into the two Young
sectors $(1,1)$ and $(2)$.  The $(1,1)$ part is still represented in
$D_{n+2}$, while the $(2)$ part reaches one Wronskian step beyond it.
The two external terms are exactly
%\begin{equation}
\[
 E_+=\tau_{n-1;(2),\varnothing}^{[1,1]}[\psi],
 \qquad
 E_-=\tau_{n;\varnothing,(2)}^{[0,0]}[\psi^*].
%\end{equation}
\]
Hence the one-step enlargement Eq.~(\ref{one-step}) is not an ad hoc repair of determinant closure.  It is the minimal
enlargement required by the $(2)$ sectors generated by repeated shifts
in one direction.

The Pfaffian reductions of these external terms possess an additional
common structure.  Both are realizations of the local three-term
Pluecker relation
%\begin{equation}
\[
 \Delta_a\delta_{bc}
 +\Delta_b\delta_{ca}
 +\Delta_c\delta_{ab}=0,
%\end{equation}
\]
with different orientations of the same row/column geometry.  Thus
the weight grading, partition growth under differentiation, one-step
Wronskian enlargement, and Pfaffian reduction form a single coherent
hierarchy rather than a collection of separate observations.

This constitutes the principal structural result of the present work.
It does not complete the generic rotating proof of the Nakamura
conjecture.  A complete proof must still show that the remaining
minors in $D_{n+2}$ and $\widehat D_{n+3}$ combine with the factors
$\sinh^2\xi$, $\cosh^2\eta$, and the zero-order term $-2n^2$ to give
the required cancellation.  What has been achieved is a restriction
of that problem to a finite and canonically generated set of
partition-labeled minors with local Pluecker relations between them.

The exact weight relation  $n^2=(n^2-1)+1$ continues to suggest a possible role for the elementary seed $\psi$.
However, weight balance alone does not determine a genuine three-slot
differential equation, and no such closure is asserted here.  The
partition-resolved determinant hierarchy should instead be regarded
as the primary structural result; any future trilinear formulation
would have to reproduce this hierarchy and its local Pluecker cells.

A natural next step is therefore to search for a closed general-$n$
identity on this partition lattice which incorporates simultaneously
the $(1,1)$, $(2)$, and mixed sectors with the coordinate coefficients
of the Nakamura operator.  The present formulation isolates precisely
which minors such an identity must contain and which extensions are
not required.
The case $n=3$ provides the first nontrivial member in which the
empty, mixed, $(1,1)$, and $(2)$ sectors all occur simultaneously.
Indeed, for $n=2$ one has $f_2=\tau^{[1]}_1$, for which the
$(1,1)$ partition is absent, whereas $f_3=\tau^{[1]}_2$ already
contains the complete second-order partition structure.  Since the
derivative rules generating these sectors are independent of the
determinant size, the $n=3$ case may be regarded as the first complete
local cell of the partition-resolved determinant hierarchy, whose
structure persists for arbitrary $n$.

This local determinant picture also provides a natural point of
contact with the general-$n$ algebraic representations of the
TS family discussed in recent work.  Those representations
organize the exact solutions globally in $n$, whereas the present
construction identifies the local determinant geometry generated by
differentiation.  Establishing the precise correspondence between
these two descriptions may provide the missing general-$n$ Pluecker
identity required to complete the rotating Nakamura conjecture.


\begin{thebibliography}{99}
\bibitem{Ernst}
F.~J.~Ernst,
``New Formulation of the Axially Symmetric Gravitational Field Problem,''
Phys. Rev. {\bf 167}, 1175 (1968).
\bibitem{Kerr1963}
R.~P.~Kerr,
``Gravitational Field of a Spinning Mass as an Example of
Algebraically Special Metrics,''
Phys.\ Rev.\ Lett.\ {\bf 11}, 237 (1963).
\bibitem{TomimatsuSato1972}
A.~Tomimatsu and H.~Sato,
``New Exact Solution for the Gravitational Field of a Spinning Mass,''
Phys. Rev. Lett. {\bf 29}, 1344--1345 (1972);
doi:10.1103/PhysRevLett.29.1344.
 \bibitem{TomimatsuSato}
A.~Tomimatsu and H.~Sato,
``New exact solution for the gravitational field of a spinning mass,''
Prog.\ Theor.\ Phys.\ {\bf 50}, 95 (1973).

\bibitem{YamazakiHori}
M.~Yamazaki and S.~Hori,
``Generalization of the Tomimatsu--Sato Solutions,''
Prog. Theor. Phys. {\bf 57}, 696--697 (1977).
\bibitem{Hori1978}
S.~Hori,
``On the Exact Solution of Tomimatsu--Sato Family for an Arbitrary
Integral Value of the Deformation Parameter,''
Prog. Theor. Phys. {\bf 59}, 1870--1891 (1978);
Erratum: Prog. Theor. Phys. {\bf 61}, 365 (1979). 
\bibitem{Vein}
P.~R.~Vein,
``Persymmetric determinants: The derivatives of determinants
with Appell function elements,''
Linear Multilinear Algebra {\bf 11}, 253--265 (1982).


\bibitem{Melikyan2025}
A.~R.~Melikyan,
``On the Yamazaki--Hori solution of the Ernst equation,''
Phys. Lett. B {\bf 868}, 139735 (2025).

\bibitem{Melikyan2026}
A.~R.~Melikyan,
``Algebraic structure of the Yamazaki--Hori solutions for the
Ernst equation,''
J. Phys. A: Math. Theor. {\bf 59}, 285205 (2026).
\bibitem{Cosgrove}
C.~M.~Cosgrove,
``New family of exact stationary axisymmetric gravitational fields
generalising the Tomimatsu--Sato solutions,''
J. Phys. A: Math. Gen. {\bf 10}, 1481--1524 (1977).
\bibitem{Weyl1917}
H.~Weyl,
``Zur Gravitationstheorie,''
Ann.\ Phys.\ {\bf 54}, 117--145 (1917).
\bibitem{Toda1967}
M.~Toda,
``Vibration of a Chain with Nonlinear Interaction,''
J. Phys. Soc. Jpn. {\bf 22}, 431 (1967).

\bibitem{Nakamura}
A.~Nakamura,
``Relation between Tomimatsu--Sato black holes and semi-infinite
Toda molecule,''
J. Phys. Soc. Jpn. \textbf{62}, 368 (1993).
\bibitem{NakamuraOhta}
A.~Nakamura and Y.~Ohta,
``Bilinear, Pfaffian and Legendre Function Structures of the
Tomimatsu--Sato Solutions of the Ernst Equation in General Relativity,''
J. Phys. Soc. Jpn. {\bf 60}, 1835 (1991).
\bibitem{FKY}
T.~Fukuyama, K.~Kamimura and S.~Yu,
``Toda lattice and Tomimatsu--Sato solutions,''
J.\ Phys.\ Soc.\ Jpn.\ {\bf 64}, 3201 (1995).


\bibitem{ImaiFukuyama}
T.~Imai and T.~Fukuyama,
``Aitken Acceleration and Stationary Axially Symmetric Solution
of Einstein Equation,''
J. Phys. Soc. Jpn. {\bf 64}, 3682 (1995).

\bibitem{FK2011}
T.~Fukuyama and K.~Koizumi,
``Toda Molecule and Tomimatsu--Sato Solution:
Towards the Complete Proof of Nakamura's Conjecture,''
J.\ Phys.\ A: Math.\ Theor.\ {\bf 44}, 345201 (2011).





\bibitem{Hirota}
R.~Hirota,
``Exact solution of the Korteweg--de Vries equation for multiple
collisions of solitons,''
Phys.\ Rev.\ Lett.\ {\bf 27}, 1192 (1971).

\bibitem{HirotaBook}
R.~Hirota,
{\it The Direct Method in Soliton Theory},
Cambridge University Press, Cambridge (2004).

\bibitem{GRH}
B.~Grammaticos, A.~Ramani and J.~Hietarinta,
``Multilinear operators: the natural extension of Hirota's
bilinear formalism,''
Phys.\ Lett.\ A {\bf 190}, 65 (1994).

\bibitem{YTSF1}
S.~Yu, K.~Toda, N.~Sasa and T.~Fukuyama,
% Bibliographic details should be checked against the original paper
% before submission.
``N-soliton solutions to the Bogoyavlenskii--Schiff equation
and a quest for the soliton solution in $(3+1)$ dimensions,''
J.\ Phys.\ A: Math.\ Gen.\ {\bf 31}, 3337 (1998).
\bibitem{Sato1981}
M.~Sato,
``Soliton equations as dynamical systems on an infinite dimensional
Grassmann manifold,''
RIMS K\^oky\^uroku \textbf{439}, 30 (1981).
\bibitem{MiwaJimboDate}
T.~Miwa, M.~Jimbo and E.~Date,
\textit{Solitons: Differential Equations, Symmetries and
Infinite Dimensional Algebras}
(Cambridge University Press, Cambridge, 2000).




\end{thebibliography}
\end{document}